\documentclass[a4paper]{article}

\usepackage[utf8]{inputenc}
\usepackage[T1]{fontenc}

\usepackage[UKenglish]{babel}

\usepackage[style=chem-acs, date=year, isbn=false]{biblatex}
\usepackage{setspace}
\usepackage[margin=25mm]{geometry}
\usepackage[version=4]{mhchem}
\usepackage{siunitx, threeparttable, graphicx, color, multirow,
  booktabs, titlesec}

\usepackage[proportional,scaled=1]{erewhon}
\usepackage[erewhon,vvarbb,bigdelims]{newtxmath}
\usepackage[defaultsans]{droidsans}

\usepackage{microtype, xcolor}

\titleformat{\section}{\bfseries\large}{}{0em}{}

\title{\sffamily Hydrogen-bond-mediated structural variation\\ of metal guanidinium formate hybrid perovskites\\ with unit cell volume}

\author{
  Zhengqiang Yang\footnote{School of Physics and Astronomy, Queen
    Mary University of London, London E1 4NS, U.K.} \and
  Guanqun Cai\footnotemark[1] \and
  Craig L. Bull\footnote{ISIS Neutron and Muon Source, Rutherford Appleton Laboratory, Chilton, Didcot, Oxon OX11 0QX, U.K.} \and
  Matthew G. Tucker\footnotemark[2]\;\textsuperscript{, }\footnote{Spallation Neutron Source, Oak Ridge National Laboratory, Oak Ridge, TN, USA 37830-6286} \and
  Martin T. Dove\footnotemark[1] \and
  Alexandra Friedrich\footnote{Institut für Geowissenschaften, Goethe-Universität Frankfurt, Altenhöferallee 1, 60438 Frankfurt am Main, Germany}\;\textsuperscript{, }
\footnote{Current address: Institut für Anorganische Chemie, Julius-Maximilians-Universität Würzburg, Am Hubland, 97074 Würzburg, Germany} \and
  Anthony E. Phillips\footnotemark[1]\;\textsuperscript{, }\footnote{\texttt{a.e.phillips@qmul.ac.uk}}}

\begin{document}
\maketitle

\onehalfspacing

\begin{abstract}
  The hybrid perovskites are coordination frameworks with the same
  topology as the inorganic perovskites, but with properties driven by
  different chemistry, including host-framework hydrogen bonding. Like
  the inorganic perovskites, these materials exhibit many different
  phases, including structures with potentially exploitable
  functionality. However, far less is known about their behaviour
  under pressure. We have studied the structures of of manganese and
  cobalt guanidinium formate under pressure using single-crystal X-ray
  and powder neutron diffraction. Remarkably, when pressure
  \emph{reduces} these materials’ volume, they transform to a phase
  isostructural to cadmium guanidinium formate, which has an
  \emph{larger} volume. Using DFT calculations, we show that this
  counterintuitive behaviour depends on the hydrogen-bonded network of
  guanidinium ions, which act as struts protecting the metal formate
  framework against compression. Our results demonstrate more
  generally that engineering desirable crystal structures in the
  hybrid perovskites will depend on achieving suitable host-guest
  hydrogen-bonding geometries.
\end{abstract}

\section{Introduction}

The hybrid perovskites are a family of materials analogous in
structure to the inorganic perovskites. In both the inorganic and
hybrid materials, ``\textbf{B} site'' cations are linked by anions
into a cubic network, with ``\textbf{A} site'' cations occupying the
cubic interstices. In the hybrid materials, however, a relatively
large linker anion such as iodide, cyanide, or formate expands the
network compared to the inorganic analogues, allowing the interstitial
\textbf{A} site to be occupied by a polyatomic organic ion such as an
alkylammonium, guanidinium, formamidinium, acetamidinium, or
imidazolium. Like their inorganic analogues, the hybrid materials
exhibit both a great diversity of potential compositions, with
hundreds of these materials reported over the past
decade,\autocite{li_chemically_2017, kieslich_same_2017} and important
functionality, most famously including solar energy
conversion\autocite{kanemitsu_photophysics_2018} but also
ferroelectric\autocite{shi_structural_2017} and caloric
behaviour.\autocite{bermudez-garcia_giant_2017}

There is every reason to expect the phase diagrams of this family of
materials to be as rich as their inorganic
counterparts.\autocite{xu_structural_2016} Indeed, because the
polyatomic linker anions lend the frameworks greater intrinsic
flexibility, we might anticipate an even greater diversity of phases
in the hybrid materials. This phase transition behaviour will depend
on fundamentally new physics and chemistry. In contrast with the
inorganic perovskites, organic \textbf{A}-site cations have a shape:
more formally, they may have intrinsic electric dipole or higher-order
multipole moments,\autocite{evans_control_2016} which will strongly
influence their structure and properties. Similarly, hydrogen bonding
between the organic guest cation and anionic framework may
dramatically change the relative stability of different
structures. These effects have only recently begun to be explored and
remain poorly understood. Yet mapping and understanding phase
transitions in the hybrid perovskites is both of intrinsic interest
from a crystal engineering perspective and of great value for
potential applications, as a means to tune these materials' electrical
and magnetic properties.

The best-explored variable in the phase diagrams of the hybrid
perovskites is temperature, with many phase changes with respect to
temperature now known.\autocite{shi_structural_2017} On the other
hand, with the exception of the well-studied lead halide perovskite
semiconductors,\autocite{postorino_pressure-induced_2017} relatively
few structural studies of materials in this family under applied
pressure have been reported,\autocite{collings_neon-bearing_2016,
  collings_structural_2016, feng_high_2016} although in some cases
vibrational spectroscopy has intriguingly indicated structural
changes.\autocite{gomez-aguirre_room-temperature_2015,
  xin_growth_2016} In particular, spectroscopic methods have revealed
high-pressure changes in many metal formate hybrid
perovskites.\autocite{maczka_raman_2014, maczka_temperature-_2014,
  maczka_comparative_2015, maczka_raman_2016,
  maczka_temperature-_2016, maczka_structural_2016}

We report here single-crystal synchrotron X-ray diffraction, powder
neutron diffraction, and density-functional theory calculations on the
metal guanidinium formate hybrid perovskites under pressure. Our
results demonstrate that the phase diagrams of these materials are
dictated by hydrogen-bonding interactions between the guanidinium and
formate ions, with implications for crystal engineering of the hybrid
perovskites more generally.

\section{Target materials}

\begin{figure}[t]
  \centering
  \includegraphics{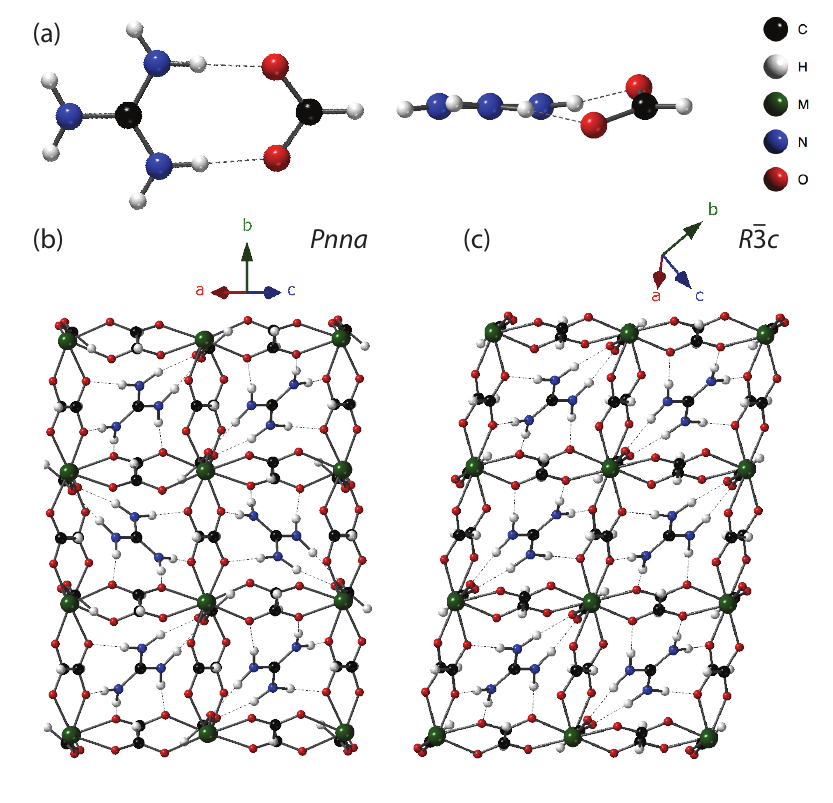}
  \caption{(a) In the crystal structures discussed here, the
    guanidinium (\ce{N(CH2)3+}) and formate (\ce{HCO2-}) ions have a
    snug hydrogen-bonded fit. 
    Depending on the metal ion, two metal guanidinium formate
    perovskite structures are known at ambient pressure: (b)~an
    orthorhombic phase in which guanidinium ions lie in two
    differently oriented planes, alternating down each column; and
    (c)~a rhombohedral phase (referred here to hexagonal axes) in
    which all guanidinium ions lie in parallel planes.}
  \label{fig:structure}
\end{figure}

We focus here on the metal guanidinium formates,
\ce{C(NH2)3[\textbf{M}^{II}(HCO2)3]}, which we henceforth abbreviate
\textbf{MGF}.  In these materials, the metal ions \textbf{M} are
linked by formate ions into a network, with the guanidinium ions
occupying the cubic interstices.\autocite{hu_metalorganic_2009} The
guanidinium ions act as struts that support the framework through the
snug hydrogen-bonded fit between guanidinium and formate ions
(Fig.~\ref{fig:structure}a). As a result of this strong interaction,
the guanidinium ions are crystallographically ordered, in contrast
with, for instance, the dimethylammonium metal
formates\autocite{jain_multiferroic_2009} and the guanidinium metal
cyanides,\autocite{xu_controlling_2016} where the guest-framework
interaction is weaker and the guest ions are disordered at room
temperature.

The materials in this family with
$\mathbf{M} = \ce{Mn}, \ce{Fe}, \ce{Co}, \ce{Ni}, \ce{Zn}$ have an
orthorhombic structure (space group $Pnna$), in which the pseudocubic
perovskite cell is distorted slightly along the face diagonal. In this
structure, alternate planes of guanidinium ions have different
orientations, forming a herringbone pattern
(Fig.~\ref{fig:structure}b). By contrast, \textbf{CdGF} adopts a
rhombohedral structure (space group $R\bar{3}c$) in which the
pseudocubic cell is distorted along the body
diagonal.\autocite{collings_compositional_2015} In this form, each
guanidinium ion has the same orientation (Fig.~\ref{fig:structure}c).

\begin{figure}[tp]
  \centering
  \includegraphics{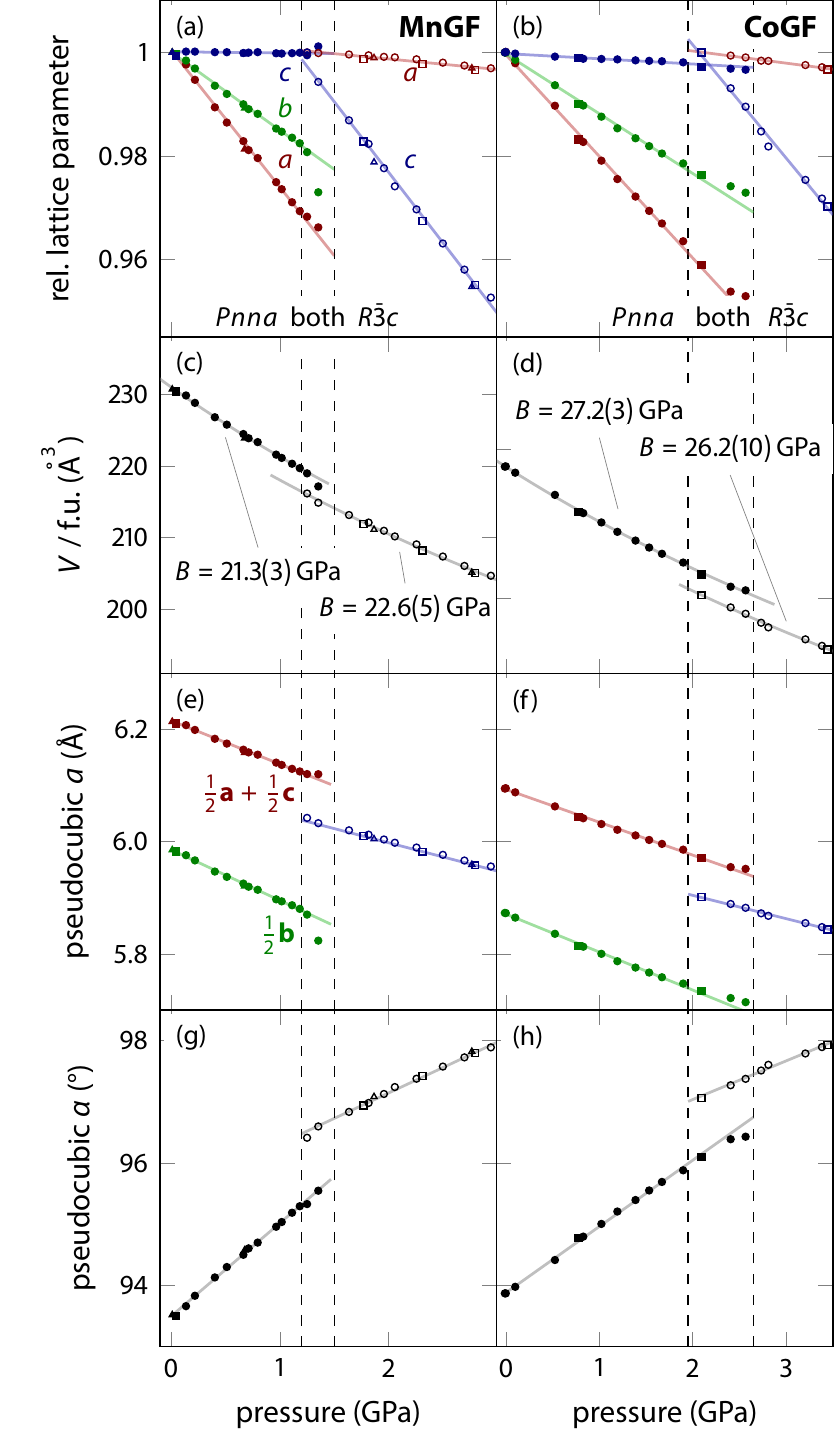}
  \caption{Crystallographic unit cell parameters of (a) \textbf{MnGF}
    and (b) \textbf{CoGF}, relative to the lowest-pressure values in
    each phase, as determined from powder neutron
    diffraction. Normalised crystallographic unit cell volumes along
    with fitted bulk moduli (see ESI), are shown in (c, d).  The same
    data can be alternatively visualised in terms of the pseudocubic
    perovskite (e, f) cell length $a$ and (g, h) lattice angle
    $\alpha$. Closed symbols represent the orthorhombic phase, open
    symbols the rhombohedral one; different symbols (circle, triangle,
    square) correspond to different sample loadings.}
    \label{fig:cell}
  \end{figure}

  We have studied the behaviour of \textbf{MnGF} and \textbf{CoGF}
  under pressure, using single-crystal laboratory and synchrotron
  X-ray and powder neutron diffraction.  We found that both
  \textbf{MnGF} and \textbf{CoGF} undergo a first-order transition
  from the ambient orthorhombic phase to a rhombohedral phase
  isostructural with \textbf{CdGF} at moderate pressures, with the two
  phases in each case coexisting over a small pressure range
  (\textbf{MnGF}: \SIrange{1.2}{1.5}{GPa}; \textbf{CoGF}:
  \SIrange{2.0}{2.6}{GPa}).  Unlike the related metal ammonium
  formates in argon pressure-transmitting medium, no indication of the
  medium entering the framework was
  observed.\autocite{collings_neon-bearing_2016} In single-crystal
  measurements, the high-pressure phase exists as a non-merohedral
  twin, with two components corresponding to the two orientations of
  the guanidinium ions in the ambient-pressure herringbone pattern
  (see ESI). Indeed, parallel twin domains are clearly visible in the
  high-pressure phase (Fig. S1). Taking layers of the two different
  guanidinium orientations to represent ``spin up'' and ``spin down'',
  the system is thus analogous to a one-dimensional Ising spin-chain:
  initially antiferromagnetic, applying pressure causes the
  nearest-neighbor interactions to become ferromagnetic, and hence
  domains of aligned guanidinium ions grow to macroscopic sizes.

Here we will first discuss the behaviour within each phase and then
consider the reasons for the phase transition itself.

\section{Strain}

It is instructive to examine the structural variation within each
phase in two different ways. First, we can simply plot the relative
change of each lattice parameter on applying pressure
(Fig.~\ref{fig:cell}a, b).  In both the orthorhombic and the
rhombohedral phases, the linear compressibility varies substantially
between the crystallographic axes (Table~\ref{tab:linear}). At the
most extreme example, in the orthorhombic phase of \textbf{MnGF}, the
linear compressibility along the $a$ axis is substantial while that
along the $c$ axis is within experimental error of zero.  This
behaviour is readily understandable in terms of the orientation of the
guanidinium ions, which act as struts, keeping the framework
relatively rigid within their plane (see Figure 1). In the
orthorhombic phase, the $c$ axis runs parallel to the plane of every
guanidinium ion, while the $a$ and $b$ axes are angled away from these
planes; thus the linear compressibility is far greater along the $a$
or $b$ axes than along $c$. By contrast, in the rhombohedral phase the
guanidinium ions lie in the $ab$ plane, and the linear compressibility
is hence greater along $c$ than along $a$ or $b$. The net effect in
both materials is that the two phases have comparable bulk moduli
(Fig.~\ref{fig:cell}c, d).

  \begin{table}
  \centering\footnotesize
  \begin{tabular}{lSS}
    \toprule
    Axis & {\textbf{MnGF}} & {\textbf{CoGF}} \\
    \midrule
    \multicolumn{3}{l}{\textbf{Orthorhombic}} \\
    $a$ & 26.7(4) & 19.5(3) \\
    $b$ & 15.2(2) & 11.4(2) \\
    $c$ & 0.03(16) & 1.10(8) \\
    \midrule
    \multicolumn{3}{l}{\textbf{Rhombohedral (hexagonal axes)}} \\
    $a=b$ & 2.04(15) & 2.34(10) \\
    $c$ & 27.3(5) & 21.9(7) \\
    \bottomrule
  \end{tabular}
  \caption{Linear compressibilities $-\partial\ell/\ell\partial P$
    (\si{TPa^{-1}}) of the target materials in the orthorhombic and
    rhombohedral phases, estimated from straight-line fits to the
    crystallographic data shown in Fig.~\ref{fig:cell}.}
  \label{tab:linear}
\end{table}

A second way to examine these data is to transform the lattice
parameters to a pseudocubic cell corresponding to the cubic perovskite
aristotype. In the orthorhombic phase, this pseudocubic cell has two
independent cell lengths and one variable angle (with the other two
fixed at \ang{90}); in the rhombohedral phase, the pseudocubic cell's
three lengths and three angles are respectively identical.  Analysing
the data in this fashion shows by contrast that the pseudocubic cell
lengths $a$ decrease in both phases of both materials at an
approximately constant rate (Fig.~\ref{fig:cell}e, f), while the pseudocubic cell
angle $\alpha$ increases (Fig.~\ref{fig:cell}g, h). In each case this reflects the
compression and collapse that would be expected of a topologically
cubic framework under pressure.

Of course, these two analyses contain exactly the same information,
but they highlight different aspects. Whilst applying pressure causes
the cubic metal formate framework to collapse, the guest guanidinium
ions act as relatively incompressible struts.

\section{Phase transition}

We now turn to the phase transition itself. At first glance it seems
puzzling that, starting from the orthorhombic phase of \textbf{MnGF}
or \textbf{CoGF}, either \emph{increasing} the unit cell volume by
replacing manganese (crystal radius $r = \SI{0.97}{\angstrom}$) or
cobalt ($r = \SI{0.885}{\angstrom}$) by cadmium ($r =
\SI{1.09}{\angstrom}$),\autocite{shannon_revised_1976} or
\emph{decreasing} the unit cell volume by applying pressure should
result in the same rhombohedral structure. 

\begin{figure}
  \centering
  \includegraphics{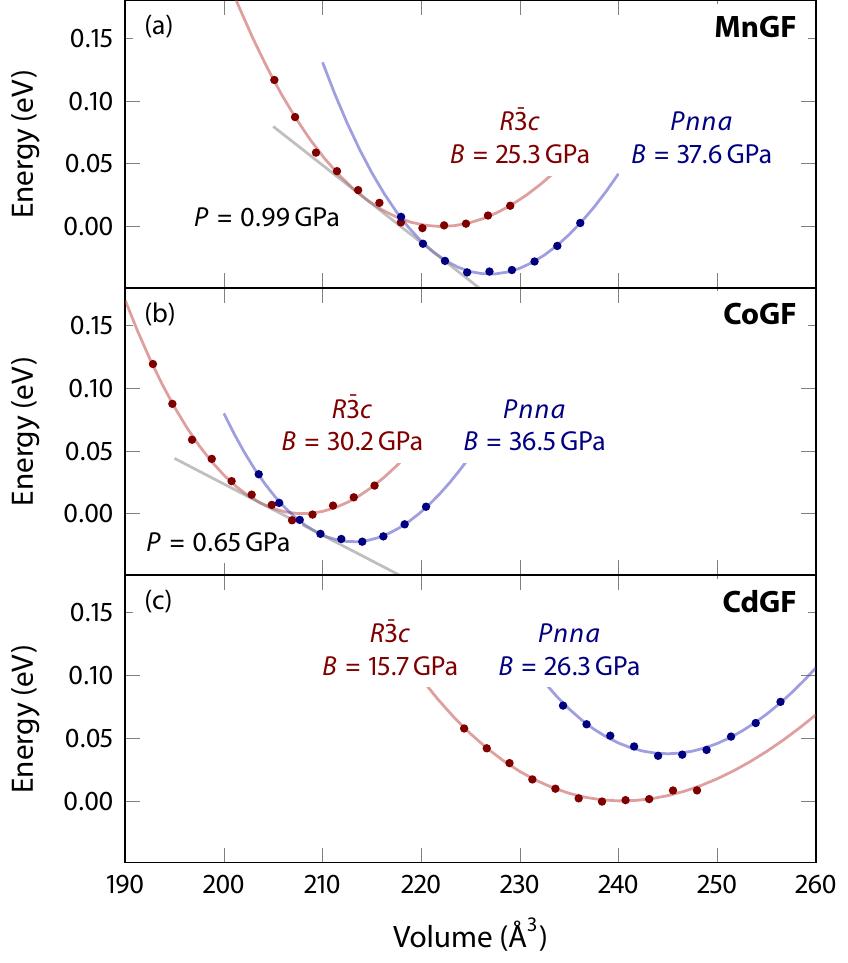}
  \caption{Lattice energy as a function of volume of the rhombohedral
    and orthorhombic phases from DFT geometry optimisations at
    constant volume, with fits to the second-order Birch-Murnaghan
    equation of state, for (a) \textbf{MnGF}, (b) \textbf{CoGF}, and
    (c) \textbf{CdGF}. Both energy and volume are given per formula
    unit. The fitted bulk modulus $B$ is labelled on the graph; full
    fitted parameters are given as ESI.}
  \label{fig:dft}
\end{figure}

To help resolve this puzzle, we used density-functional theory
calculations to calculate the energy of each phase as a function of
volume for \textbf{MnGF}, \textbf{CoGF} and \textbf{CdGF}.  In
agreement with the experimental data, our results show that in
\textbf{MnGF} and \textbf{CoGF}, the orthorhombic phase is the most
stable at zero pressure, while the rhombohedral phase, with smaller
volume, is favored at higher pressures (Fig.~\ref{fig:dft}a, b). The predicted
transition pressures are \SI{0.99}{GPa} (\textbf{MnGF}) and
\SI{0.65}{GPa} (\textbf{CoGF}). It is not surprising that these values
are both smaller than observed experimentally: this first-order phase
transition involves substantial rearrangement of the guanidinium ions,
and the pressure at which the two phases nominally have the same
enthalpy should thus be considered a lower bound rather than a
quantitative prediction of the phase transition pressure. Indeed, one
might expect that in the larger Mn cell this rearrangement should be
slightly easier than in the smaller Co analogue, rationalising the
observations that both the difference between the nominal DFT and
experimentally observed phase transition pressures, and the pressure
range where the phases coexist, are smaller for \textbf{MnGF} than for
\textbf{CoGF}.

By contrast, in \textbf{CdGF}, although the stablest orthorhombic
structure again has greater volume than the stablest rhombohedral
structure, the rhombohedral phase is favored at all cell volumes
(Fig.~\ref{fig:dft}c). Again, this is in agreement with the experimental
observation that no orthorhombic phase has been observed in this
material.

\section{Discussion}

The difference between these materials' behaviour can be rationalised
in terms of the hydrogen bonding between guanidinium and formate
ions. Table~\ref{tab:NO-dist} shows the \ce{N-H\bond{...}O} distance
in \textbf{MnGF} and \textbf{CdGF}, as determined from single-crystal
X-ray diffraction experiments and DFT calculations. In each case the
DFT values are \SIrange{0.05}{0.08}{\angstrom} smaller than the
diffraction results at ambient temperature and pressure. In
\textbf{MnGF}, the DFT \ce{N\bond{...}O} distances fall by at most
\SI{0.025}{\angstrom} across the phase transition. Even considering
the energy minima, rather than the structures immediately before and
after the phase transition, the differences in distance between the
phases range from \SIrange{0.03}{0.04}{\angstrom}. This result is
slightly smaller than the experimental values of
\SIrange{0.03}{0.09}{\angstrom}, where to provide the most accurate
experimental comparison, we have combined the atomic coordinates from
single-crystal diffraction with lattice parameters from powder
diffraction at a pressure where the phases coexist.

On the other hand, in \textbf{CdGF}, the difference in the DFT
\ce{N\bond{...}O} distance between the two energy minima is
\SI{0.11}{\angstrom}, four times the corresponding value in
\textbf{MnGF}. Moreover, the absolute DFT \ce{N\bond{...}O} distance
in the putative orthorhombic phase, \SI{2.99}{\angstrom}, is
substantially larger than the DFT value from any experimentally
observed phase (\SIrange{2.87}{2.91}{\angstrom}). Thus it seems that
the hypothetical orthorhombic unit cell in \textbf{CdGF} is both too
large and too rigid to allow effective hydrogen bonding.

This is consistent with our observations of these structures'
flexibility more generally. As previously noted, the rhombohedral
structure is distorted along the pseudocubic body diagonal
(\emph{i.e.}, the hexagonal $c$ axis; see Fig.~\ref{fig:structure}c), which is
perpendicular to all guanidinium ions, and is therefore relatively
flexible along this direction. On the other hand, the herringbone
arrangement of guanidinium ions makes the orthorhombic structure more
rigid. Thus the rhombohedral structure is able to accommodate
favorable hydrogen-bonding distances in both \textbf{MnGF} and
\textbf{CdGF}; by contrast, the more rigid orthorhombic structure is
unable to distort in this way.

At this point we pause to consider the extent and nature of the
agreement between our experimental and computational data. In addition
to uncertainties associated, for instance, with the specific choice of
exchange-correlation functional, the DFT methodology used here has two
features that fundamentally differentiate it from experiment. First,
DFT does not intrinsically take thermal vibrations into account, and
thus effectively produces data at absolute zero temperature. Second,
fitting optimised energy as a function of cell volume considers only
distortions at the gamma point (that is, those in which every unit
cell distorts in the same way). The first of these points helps to
explain why, in neglecting thermal expansion, the DFT underestimates
the unit cell volume and hence overestimates the bulk modulus. The
second explains why the DFT predicts that the orthorhombic phase is
mechanically stiffer than the rhombohedral one (Fig.~\ref{fig:dft}),
while the experimental bulk moduli of the phases are similar
(Fig.~\ref{fig:cell}c, d). Indeed, the DFT isolates the specific sense
of flexibility that we argue is responsible for the phase
transformation behaviour: the ability of the metal formate framework
to accommodate the guanidinium ions at a variety of unit cell volumes;
thus the apparent discrepancy between experimental and simulated bulk
moduli does not contradict our argument above.

\begin{table*}
  \centering\footnotesize
  \begin{tabular}{lllr}
    \toprule
    \textbf{Material} & \textbf{Conditions}                                               & \textbf{Phase} & \textbf{\ce{N-H\bond{...}O} distance (Å)} \\
    \midrule
    \textbf{MnGF}     & SCXRD, \SI{0}{GPa}, \SI{293}{K}\autocite{hu_metalorganic_2009}        & $Pnna$         & 2.9529(19), 2.976(2), 2.9904(16)          \\
                      & PND, \SI{1.25}{GPa}, ambient $T$                                & $Pnna$         & 2.906(5), 2.919(5), 2.965(5)             \\
                      &                               & $R\bar{3}c$    & 2.879(10)                                \\
    \cmidrule{2-4}
                      & DFT, minimum energy                                               & $Pnna$         & 2.8964, 2.8968, 2.9136                    \\
                      & DFT, minimum energy                                               & $R\bar{3}c$    & 2.8705                                    \\
    \cmidrule{2-4}
                      & DFT, \SI{220}{\angstrom^3}                                        & $Pnna$         & 2.8580, 2.8639, 2.8749                    \\
                      & DFT, \SI{214}{\angstrom^3}                                        & $R\bar{3}c$    & 2.8498                                    \\
    \midrule
    \textbf{CdGF}     & DFT, minimum energy                                               & $Pnna$         & 2.9883, 2.9883, 2.9883                    \\ 
                      & DFT, minimum energy                                               & $R\bar{3}c$    & 2.8736                                    \\ 
    \cmidrule{2-4}
                      & SCXRD, \SI{0}{GPa}, \SI{300}{K}\autocite{collings_compositional_2015} & $R\bar{3}c$    & 2.927(3)                                  \\
    \bottomrule    
  \end{tabular}
  \caption{\ce{N-H\bond{...}O} distances, in the orthorhombic and
    rhombohedral phases of \textbf{MnGF} and \textbf{CdGF}, from
    single-crystal X-ray and powder neutron diffraction and
    DFT modelling. For the powder neutron data, the atomic
      coordinates were fixed at their value from single-crystal X-ray
      models at the nearest available pressure.}
  \label{tab:NO-dist}
\end{table*}

As a final comparison, we consider related manganese(II) formate
perovskites in which host-guest hydrogen bonding is less important. In
dimethylammonium manganese formate, under ambient conditions, the
manganese-formate framework has the same rhombohedral structure as
discussed above, with the dimethylammonium ions disordered about the
threefold axis for want of a strongly bound hydrogen-bonding
site.\autocite{jain_multiferroic_2009, duncan_local_2016} This suggests
that host-guest hydrogen bonding is not needed to stabilise the
rhombohedral phase.  An even more dramatic example is provided by the
material ``\ce{[Mn(HCO2)3].$n$H2O}'', which has no bulky
\textbf{A}-site cation at all.  Under ambient conditions, it has the
same rhombohedral structure as discussed above, with guest water
molecules occupying the cubic interstices
($a = \SI{8.327}{\angstrom}$,
$c = \SI{22.890}{\angstrom}$).\autocite{zhou_novel_2006} The original
report suggested that this compound contains manganese(III)
ions. However, the crystals were colorless and the \ce{Mn-O} bond
lengths were \SI{2.190}{\angstrom}; both of these observations suggest
that the correct oxidation state is
manganese(II),\autocite{sidey_universal_2014} with charge balance
preserved by a guest hydronium ion, \ce{[Mn(HCO2)3].H3O.$n$H2O}. (For
comparison, the closely related compound
\ce{[Mn(HCO2)3].$\tfrac12$CO2.$\tfrac14$HCOOH.$\tfrac23$H2O}, which
unambiguously contains manganese(III) ions, is dark red and has an
\ce{Mn-O} bond length of
\SI{2.001}{\angstrom}.\autocite{cornia_manganeseiii_1999}) If this
oxidation state assignment is accepted, then this material
demonstrates that the rhombohedral structure is stable even in the
absence of a bulky organic \textbf{A}-site cation.

\section{Conclusion}

In conclusion, we have identified a new high-pressure phase in the
guanidinium metal formate perovskites \textbf{MnGF} and \textbf{CoGF}
that is isostructural with the ambient-pressure structure of
\textbf{CdGF}. Our experimental and modelling data demonstrate that
the host-guest hydrogen bonding between guanidinium and formate ions
plays a crucial role in determining which phase is the most stable:
the rhombohedral structure is able to accommodate a wide range of cell
volumes, while the orthorhombic structure provides a snug fit at
optimal cell volume but cannot as easily distort. Thus the
rhombohedral structure is favored for both small (\textbf{MnGF} and
\textbf{CoGF} under pressure) and large (\textbf{CdGF}) unit cells.
Host-guest hydrogen-bonding interactions also strongly influence
distortion within each phase, with the linear compressibility being
notably smaller in directions where the guanidinium ions are able to
resist compression by acting as ``struts'' within the framework.  More
generally, our results provide a further demonstration of the complex
interplay between framework and guest in determining the structures of
the hybrid perovskites. In contrast with their inorganic analogues,
host-guest hydrogen bonding may greatly stabilise particular
structures but only over a relatively small pressure, temperature, or
composition range. The complexity of the resulting phase diagrams, and
the consequences for these materials' properties, will reward
considerable further investigation.

\section{Methods}

All samples were synthesised by literature
methods.\autocite{hu_metalorganic_2009}

In the laboratory X-ray measurements, a single crystal of
\textbf{MnGF} was loaded in paraffin oil in a diamond anvil cell at a
pressure of \SI{1.70(3)}{GPa}. For the synchrotron X-ray experiments,
single crystals were loaded in 4:1 methanol-ethanol
pressure-transmitting medium in a diamond anvil cell and full data
collections were performed on I19 (Diamond Light Source) at several
pressures up to \SI{2.14(3)}{GPa} (\textbf{MnGF}) or \SI{1.75(3)}{GPa}
(\textbf{CoGF}). In each case a small ruby sphere was included as
internal pressure reference; the pressure was determined from the ruby
fluorescence wavelength.\autocite{ruby} For the neutron measurements,
perdeuterated powder samples of both materials were loaded, again in
4:1 methanol-ethanol medium, in a Paris-Edinburgh cell with a small
piece of lead as internal pressure reference; data were collected on
PEARL (ISIS Neutron and Muon Source)\autocite{bull_pearl:_2016} while
the pressure was monitored from refinement of the lead cell parameter.

In each of the DFT measurements, the cell parameters and atomic
positions were allowed to relax while maintaining the symmetry of the
appropriate phase and the cell volume. For \textbf{MnGF} and
\textbf{CoGF} a G-type antiferromagnetic arrangement of magnetic
moments was used. Calculations were perfomed with CASTEP
18.3\autocite{CASTEP} using the PBEsol functional\autocite{PBEsol} and
Tkatchenko-Scheffler semi-empirical dispersion
correction.\autocite{SEDC-TS}

Full experimental and computational details, along with refined
crystal structures and fit parameters, are given as ESI.

\section{Acknowledgements}

We gratefully acknowledge Bj\"orn Winkler (Goethe-Universit\"at
Frankfurt) for collaboration on preliminary experiments for this work;
Nicholas Funnell (ISIS Neutron and Muon Source) and Viswanathan
Mohandoss (QMUL) for assistance with the neutron experiments; Mark
Warren and David Allan (Diamond Light Source) for assistance with the
X-ray experiments; and Keith Refson (Royal Holloway, University of
London) for helpful discussion about the DFT calculations. We are
grateful to ISIS Neutron and Muon Source and Diamond Light Source for
the award of beam time and to the UK Materials and Molecular Modelling
Hub for computational resources, which is partially funded by EPSRC
(EP/P020194/1). ZY and GC thank the Chinese Scholarships Council for
scholarships. AF acknowledges financial support from the DFG, Germany,
within priority program SPP1236 (Project No. FR-2491/2-1) and from
Goethe-Universit\"at Frankfurt. AEP thanks EPSRC for funding
(EP/L024977/1).

\newpage

\renewcommand*{\bibfont}{\small}
\singlespacing
\printbibliography

\end{document}